\documentclass[12pt]{iopart}
\usepackage[T1]{fontenc}
\usepackage[latin1]{inputenc}
\usepackage{subfigure}
\usepackage{floatflt}
\usepackage{graphicx}
\usepackage{setspace}

\makeatletter
\input epsf
\usepackage{iopams} 
\begin{document}
\begin{center} 

\title{Systematics of global event properties in p-p and d-Au collisions at $\sqrt s_{NN}$=200GeV$^{\dagger}$}

\end{center} 

\footnote[0]{$^{\dagger}$ Results presented here were presented as a poster at Quark Matter 2004. }

\author{Levente Molnar \ddag\ (for the STAR collaboration)}  

\address{\ddag\ Department of Physics, Purdue University, West Lafayette, Indiana 47907 \\ 
molnarl@physics.purdue.edu}

\begin{abstract}
The spectra of identified charged particles were studied as a function of multiplicity at midrapidity in p-p and d-Au collisions and compared to Au-Au collisions at $\sqrt s_{NN}$=200 GeV. The spectra of heavier particles in the most central d-Au collisions and in the highest multiplicity p-p collisions are harder than in peripheral Au-Au collisions. The spectra were studied within the blast wave model framework. The extracted kinetic freeze out temperature smoothly decreases, while the average flow velocity parameter increases with increasing charged multiplicity. The particle ratios of $K^{-}/\pi$ and $\bar{p}/\pi$ show moderate change with increasing charged multiplicity. The $K^{-}/\pi$ in the highest multiplicity p-p collisions is below the ratio in central Au-Au. 

\end{abstract}

	High energy p-p and d-Au collisions provide important references for heavy ion collisions. It is also suggested that they constitute a plausible environment to search for the deconfinement phase transition\cite{rolf}, which is a soft non-perturbative QCD phenomenon. Such a phase transition could occur in high multiplicity collisions, but such collisions are contaminated by hard-QCD processes. As a first step in the study, the minimum bias data sample is investigated as a function of charged hadron multiplicity. 

\section{Data Analysis}
Measurements were carried out in the STAR experiment. The charged particles were measured in the central Time Projection Chamber\cite{tpc} (TPC) within $|$$\eta$$|$ $\le$ 1.2  and in the Forward TPCs within 2.5 $\le$ $|$$\eta$$|$ $\le$ 4.0. The Au-Au data set \cite{olga} and the p-p data set are separated into centrality/multiplicity classes based on the charged particle multiplicity within $|$$\eta$$|$ $\le$ 0.5. For d-Au collisions the three centrality classes are based on the charged particle multiplicity within -3.8 $\le$$\eta$ $\le$ -2.8.
The charged particles: $\pi^{+}$, $\pi^{-}$, $K^{+}$, $K^{-}$ and $\bar{p}$, were identified in the TPC by their ionization energy loss.

\begin{figure}[h]
\begin{spacing}{0.8}
\begin{flushright}
\hfill{\includegraphics*[%
width=1.05\columnwidth]{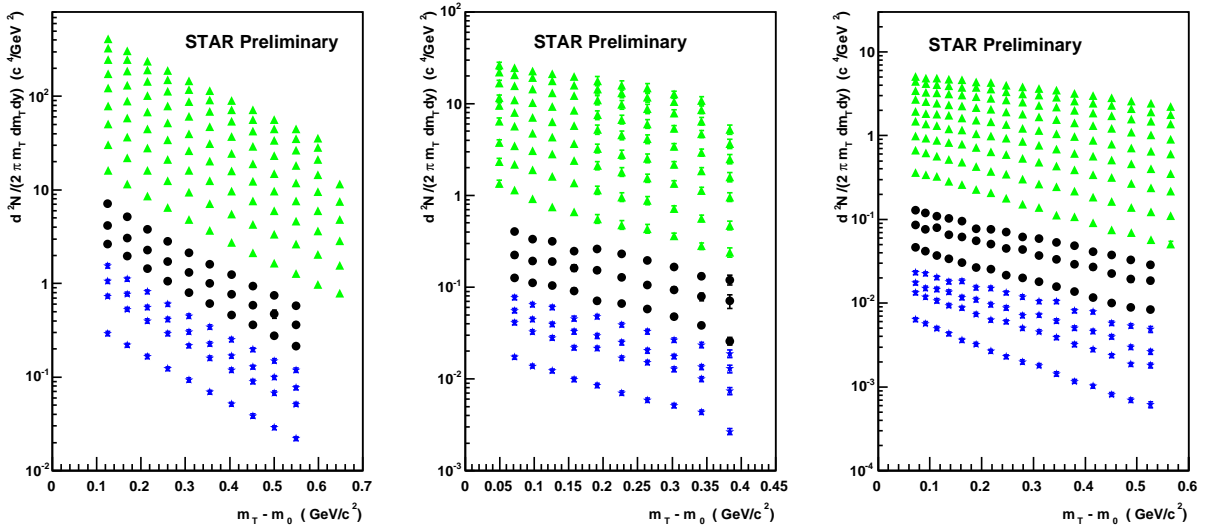}}
\end{flushright}
\caption{$m_{T}$ spectra of negative particles: $\pi^{-}$ (left), $K^{-}$ (middle) and $\bar{p}$ (right). Au-Au data\cite{olga} are shown in green triangles, d-Au data are shown in black circles and p-p data are shown in blue stars. Errors are statistical.   }
\label{tfit1}
\end{spacing}
\end{figure}
 The measured momentum was corrected for energy loss. Particle spectra were corrected for acceptance and tracking inefficiency. Pions were corrected for weak decays. In p-p and in peripheral d-Au the spectra were corrected for vertex finding inefficiency and fake vertices. More analysis details can be found in \cite{olga}.
\section{Results}

\begin{floatingfigure}[r]{0.50\columnwidth}%
\begin{spacing}{0.8}
\hspace*{-0.4in}
\includegraphics*[%
  width=0.50\columnwidth,
  keepaspectratio]{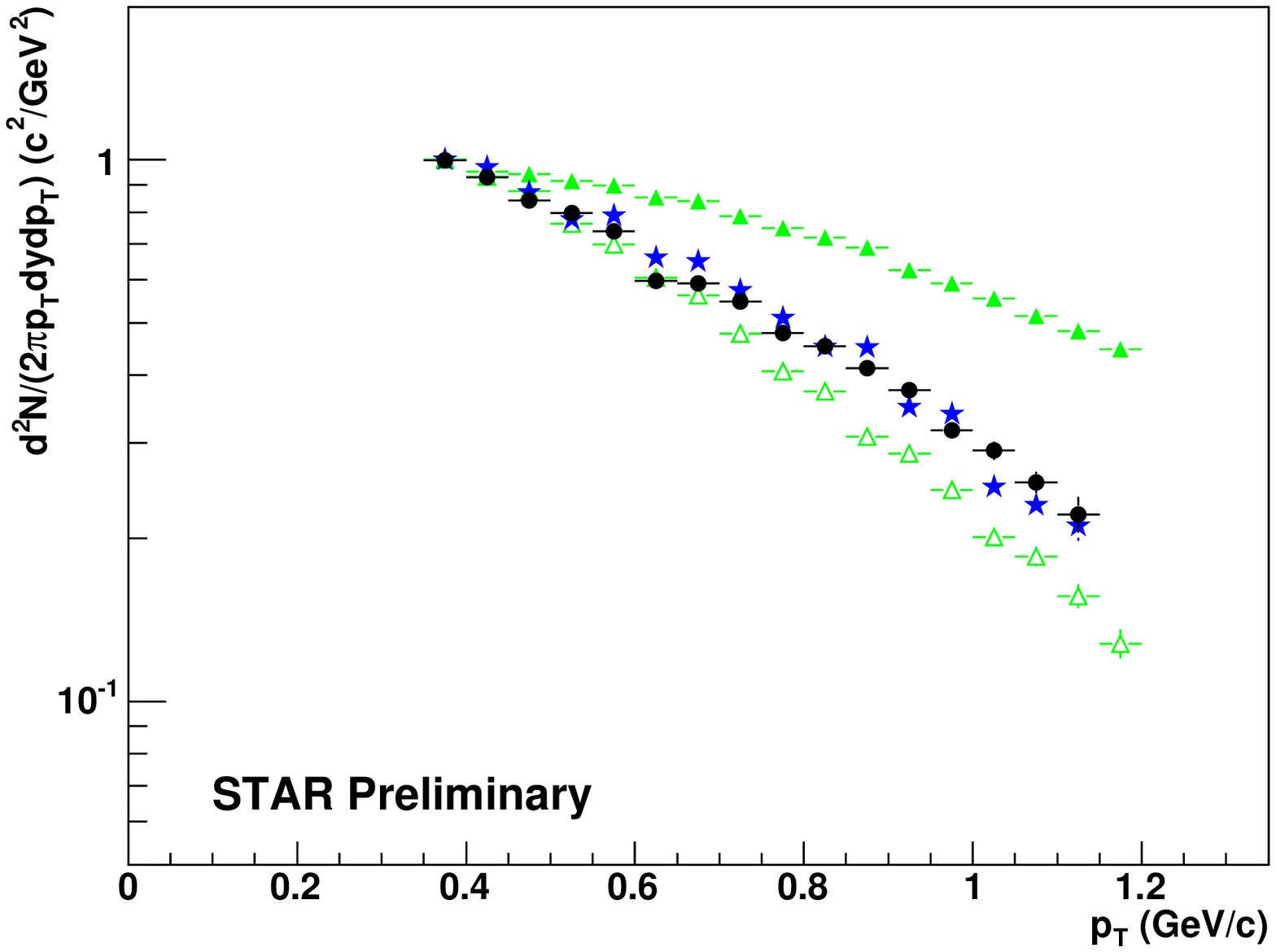}
\vspace*{-0.05in}
\hspace*{0.5in}\caption{}
\noindent \footnotesize {Antiproton spectra for Au-Au, d-Au and p-p collisions. Central Au-Au is shown in filled green triangles, peripheral Au-Au is shown in empty green triangles, central d-Au is shown in black circles, and the highest multiplicity p-p is shown in blue stars. Errors are statistical.}
\smallskip{}
\smallskip{}
\smallskip{}
\end{spacing}
\end{floatingfigure}%
Fig. 1 shows the transverse mass spectra measured in -0.3 $\le$ y $\le$ 0.3. In triangles nine centrality bins are shown for Au-Au collisions ( from top to the bottom, central to peripheral ). In circles the three centrality bin of d-Au are shown and scaled down by a factor of 2. In stars the p-p data are shown for the four multiplicity bins, scaled down by a factor of 4. For heavier particles the spectra harden with increasing multiplicity/centrality. In heavy ion collisions this hardening can be understood as an increasing radial flow from peripheral to central collisions. On the other hand the highest multiplicity p-p and the most central d-Au spectra are harder than the peripheral Au-Au spectra but not as hard as the central Au-Au as shown in Fig. 2. In p-p and d-Au collisions radial flow is presumably not present. 

One would like to investigate these systems in a common model framework. We use the blast wave model\cite{bw}, which assumes a boosted thermal source in the transverse direction. There are three parameters in the model: the kinetic freeze out temperature $T_{kin}$, the transverse flow velocity $\beta$ and the $n$ parameter that is the exponent of the flow profile: $\beta$=$\beta_{S}(r/R)^{n}$, where $\beta_{S}$ is the surface velocity. The model can describe simultaneously the $K^{+}$, $K^{-}$ and $\bar{p}$ spectra and the $p_{T}$ $\ge$ 0.5 GeV/c region of the $\pi^{+}$, $\pi^{-}$ spectra. The low momentum pions were excluded from the fit as they are significantly contaminated by resonance decays.
\begin{figure}[h]
\begin{spacing}{1.0}
\begin{flushright}
\hfill{\includegraphics*[%
width=1.05\columnwidth]{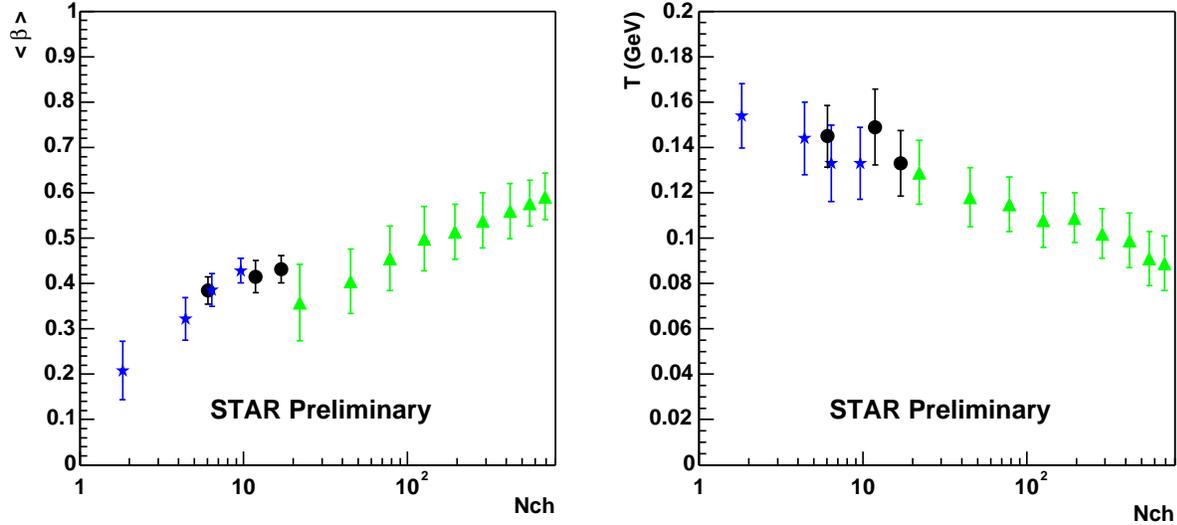}}
\end{flushright}
\caption{Average flow velocity (left panel) and kinetic freeze out temperatures (right panel) 
as a function of uncorrected charged multiplicity within $|$$\eta$$|$ $\le$ 0.5. The p-p data are shown in blue stars, d-Au data in black dots and Au-Au data in green triangles. Errors are systematic.}
\label{tfit3}
\end{spacing}
\end{figure}
Fig. 3 shows the obtained $\beta$ and $T_{kin}$. The $n$ parameters ranges between: 2.0 - 0.7 and 1.6 - 1.2 in p-p and d-Au.  There seems to be a smooth decrease in $T_{kin}$ from p-p, to d-Au to Au-Au collisions. 
The flow velocity increases with increasing charged multiplicity. The trend in p-p and d-Au seem to be different from Au-Au however the physical meaning of $\langle\beta\rangle$ may be different. For example in p-p and d-Au the increase could due to initial $k_{T}$ broadening, multiple soft interactions and increasing jet contribution.

Similar trend is observed for the average transverse momenta of antiprotons, negative kaons, which is shown in Fig. 4. The pion $\langle p_{T} \rangle$ is extracted from Bose Einstein fit and $\langle p_{T}\rangle$  of kaons and antiprotons are extracted from the blast wave fit. Pion $\langle p_{T}\rangle$  shows almost no dependence on charged multiplicity or collision system while the heavier particles $\langle p_{T}\rangle$ exhibit an increase with increasing charged multiplicity.

\begin{figure}[h]
\begin{spacing}{1.0}
\begin{flushright}
\hfill{\includegraphics*[%
width=1.05\columnwidth]{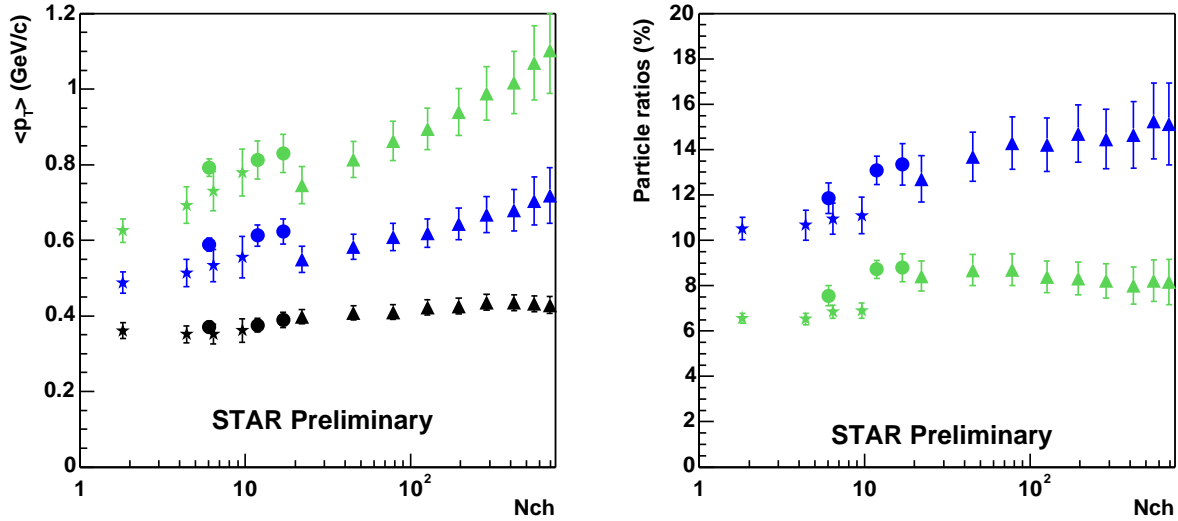}} 
\end{flushright}
\caption{Average transverse momenta (left panel) and particle ratios (right panel) as a function of uncorrected charged multiplicity within $|$$\eta$$|$ $\le$ 0.5. The p-p data are shown in stars, d-Au data in dots and Au-Au data in triangles. In the left panel:  The $\langle p_{T} \rangle$ of pions are shown in black, kaons in blue and antiprotons in green. In the right panel: $K^{-}$/$\pi$ ratio is shown in blue and $\bar{p}$/$\pi$ in green. Errors are systematic.}
\label{Ratios_and_meanpt.eps}
\end{spacing}
\end{figure}

In Fig. 4 ( right panel )  particle ratios of the integrated yields in $|$y$|$ $\le$ 0.3  are shown: $K^{-}$/$\pi$ and $\bar{p}$/$\pi$.  The ratios show a moderate change with increasing charged multiplicity. Our measurements are consistent with E735 at Tevatron\cite{rolf3,rolf2} where the multiplicity dependence of the particle ratios were also studied. The $K^{-}$/$\pi$ ratio in p-p, even at very high multiplicity as measured by E735\cite{rolf3,rolf2} appears to be smaller than central Au-Au value. It has been shown in \cite{fqw}, that the $K^{-}$/$\pi$ in heavy ion and minimum bias p-p collisions follow a systematic trend in the Bjorken estimate of initial energy density. Furthermore Scavenius et al. \cite{dumitru} argue that a larger $K$/$\pi$ ratio in Au-Au than in high-multiplcity p-p at similar energy density may indicate an onset of a deconfined phase.

\section{Conclusions}
In summary, identified $\pi$, K, and $\bar{p}$ spectra are measured in mid-rapidity and studied as a function of event multiplicity. Hardening of the particle spectra with increasing multiplicity/centrality was observed for p-p and d-Au. The data are studied within the blast wave model. The extracted kinetic freeze out temperature smoothly decreases with increasing charged multiplicity, while  the extracted $\langle\beta\rangle$ increases. The particle ratios do not show significant change.

\end{document}